\documentclass[twocolumn,superscriptaddress,PRL,nofootinbib,letterpaper]{revtex4-2}
\usepackage[english]{babel}
\usepackage{amsmath}
\usepackage{amsfonts}
\usepackage{amssymb}
\usepackage{array} 
\usepackage{dcolumn}
\usepackage{longtable}
\usepackage{hyperref}
\usepackage{float}
\usepackage{bbm}
\usepackage{bm}
\usepackage{graphicx}
\usepackage{color}
\usepackage{comment}

\definecolor{RevisionColor}{rgb}{0.325, 0.58, 0.722}

\definecolor{JulienColor}{rgb}{0.95, 0.55, 0.05}

\begin{document}
\title{Transient contacts between filaments impart its elasticity to branched actin}
\author{Mehdi Bouzid$^\S$}
\affiliation{Université Paris-Saclay, CNRS, LPTMS, 91405, Orsay, France}
\affiliation{Universit\'e Grenoble Alpes, CNRS, 3SR, 38400 Saint Martin d’Hères, France}
\author{Cesar Valencia Gallardo$^\S$}
\affiliation{PMMH, CNRS, ESPCI Paris, PSL University, Sorbonne Université, Université Paris-Cité, F-75005, Paris, France}
\author{Magdalena Kopec} 
\affiliation{PMMH, CNRS, ESPCI Paris, PSL University, Sorbonne Université, Université Paris-Cité, F-75005, Paris, France}
\author{Lara Koehler}
\affiliation{Université Paris-Saclay, CNRS, LPTMS, 91405, Orsay, France}
\affiliation{Max Planck Institute for the Physics of Complex Systems, Dresden, Germany}
\author{Giuseppe Foffi}
\affiliation{Université Paris-Saclay, CNRS, LPTMS, 91405, Orsay, France}
\author{Olivia du~Roure}
\email{olivia.duroure@espci.fr}
\affiliation{PMMH, CNRS, ESPCI Paris, PSL University, Sorbonne Université, Université Paris-Cité, F-75005, Paris, France}
\author{Julien Heuvingh}
\email{julien.heuvingh@espci.fr}
\affiliation{PMMH, CNRS, ESPCI Paris, PSL University, Sorbonne Université, Université Paris-Cité, F-75005, Paris, France}
\author{Martin Lenz}
\email{martin.lenz@universite-paris-saclay.fr}
\affiliation{Université Paris-Saclay, CNRS, LPTMS, 91405, Orsay, France}
\affiliation{PMMH, CNRS, ESPCI Paris, PSL University, Sorbonne Université, Université Paris-Cité, F-75005, Paris, France}

\def\thefootnote{\S}\footnotetext{These authors contributed equally to this work.}

\begin{abstract}
Branched actin networks exert pushing forces in eukaryotic cells, and adapt their stiffness to their environment.
The physical basis for their mechanics and adaptability is {however} not understood{. Indeed, here we show that} their high density and low connectivity place them outside the scope of standard elastic network models for actin. 
We combine high-precision mechanical experiments, molecular dynamics simulations and a mean-field elastic theory to show that they are instead dominated by the proliferation of interfilament contacts under compression.
This places branched actin in the same category as undercoordinated, fibrous materials such as sheep's wool.
When the network is grown under force, filaments entangle as if knitted together and trap contacts in their structure. Trapped contacts play a similar role as crosslinkers in rigidifying the network, and are thus key to its active adaptive mechanics.
\end{abstract}

\maketitle
Cells dynamically assemble actin into branched networks to push on their environment during cell migration, endocytosis or the healing of cell ruptures~\cite{Blanchoin:2014}. These networks are among the stiffest found in the cytoskeleton, and can generate large pushing stresses against stationary obstacles~\cite{Marcy:2004,Chaudhuri:2007,Pujol:2012,Bauer:2017}. When grown against a controlled opposing stress in an atomic force microscope (AFM)~\cite{Bieling:2016} or inside the cell~\cite{Mueller:2017}, branched networks alter their own architecture to adapt to their mechanical environment, although the underpinnings of this adaptability is unclear.

Microscopically, individual filaments in branched networks sequentially incorporate monomers at their barbed ends.
Concurrently, Arp2/3 protein complexes bind to their sides and nucleate new secondary branches, endowing the network with a tree-like topology\cite{Mullins98}. Several biochemical mechanisms prevent the uncontrolled proliferation of new branches, including the localized activation of Arp2/3 complexes at the outer surface of the network and the constant inactivation of growing barbed ends by capping proteins\cite{Loisel1999}. Branched network growth is further promoted by the presence of Rho-family GTPases and other nucleation-promoting factors (NPFs) at the cell membrane~\cite{Lappalainen2022}.

\begin{figure*}[t]
\centering
\includegraphics[width=\textwidth]{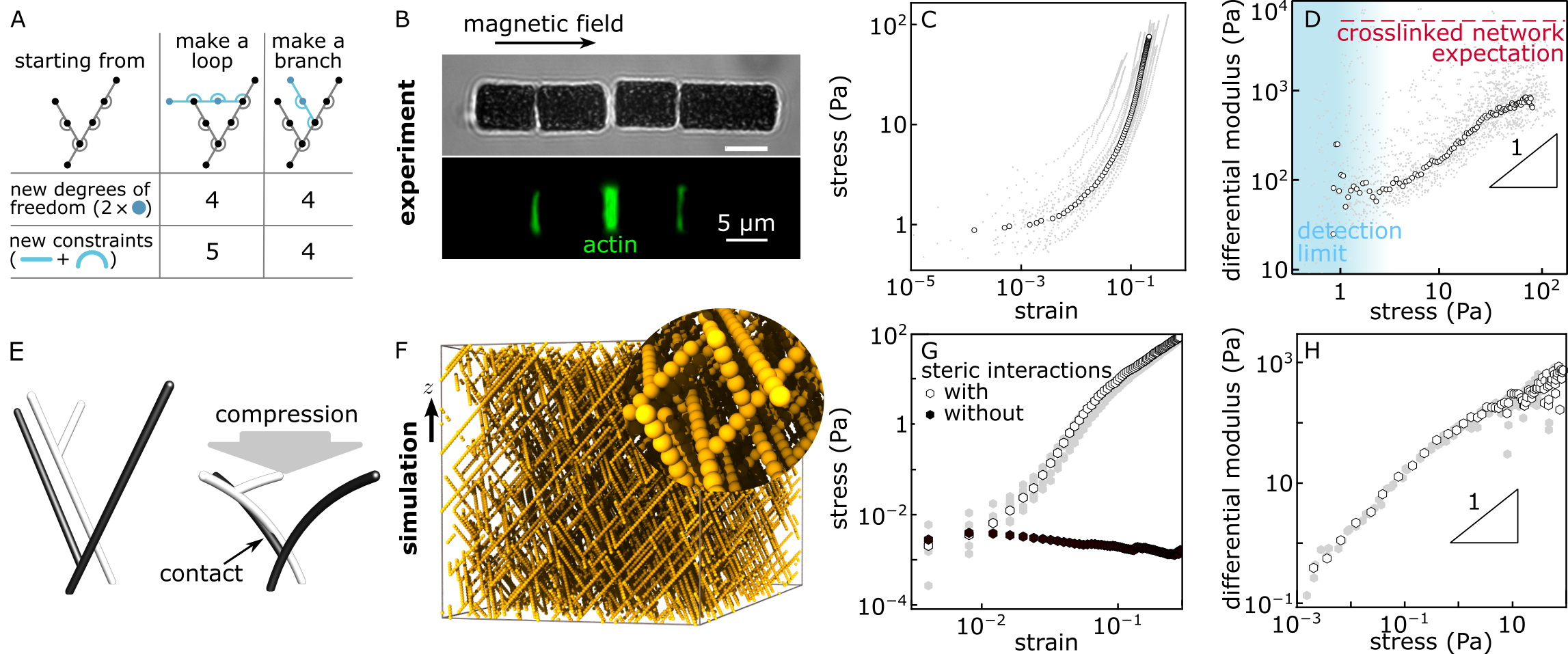}
\caption{\label{fig:basics}
\textbf{Tree-like branched networks have low connectivity, resulting in a vanishing linear elastic response.}
(\textbf{A})~In contrast to the formation of a loop, the addition of a branch does not help stabilize a tree-like network. Networks with loops (\emph{central column}), including crosslinked networks, are thus typically rigid due to their having more constraints than degrees of freedom. Loopless branched networks (\emph{right column}), on the other hand, are at most critically connected. {In both cases we add two vertices (pictured by circles)}, each of which brings 2 degrees of freedom in 2D. Line segments materialize spring-like constraints and circular arcs denote constraints of fixed angle.
(\textbf{B})~Microcylinder-based setup: chain of magnetic microcylinders (\emph{top:}~Bright-field image) between which actin has polymerized (\emph{bottom:}~fluorescence image of green-labelled actin).
(\textbf{C})~Stress-strain curves (\emph{grey}) and average (\emph{open symbols}) for branched networks grown under a weak magnetic field of $B=3\,\text{mT}$, implying a small growth stress $< 3\,\text{Pa}$.
{Data obtained by probing $N=19$ independent networks.}
(\textbf{D})~Corresponding stiffness-strain curve. It is unclear whether the elastic modulus converges to a finite value at small stress, and if it does that value is smaller than $100\,\text{Pa}$. {The red dashed line materializes the $6\,\textrm{kPa}$ small-stress modulus expected for a crosslinked network of comparable density.}
(\textbf{E})~We propose that in branched networks, the formation of contacts upon compression rigidifies the network as additional constraints would.
(\textbf{F})~Randomly generated branched network used in our simulations. The filaments are unstressed and devoid of contacts by construction, as if grown under vanishing stress.
(\textbf{G})~Average stress-strain curves with and without inter-filament interactions {($N=6$)}. 
(\textbf{H})~corresponding stiffness-strain curves. Throughout the paper, simulations reveal experimentally inaccessible small-stress regimes. Here they confirm the absence of an elastic plateau at low stress. 
}
\end{figure*}

While a direct dependence of such biochemical processes on an applied force contributes to the adaptability of branched actin subjected to large stresses~\cite{Risca:2012,Li:2022}, 
little is known about the possible role of other, purely physical modifications in encoding its mechanical history. Our physical understanding of actin elasticity indeed largely derives from models of crosslinked elastic networks~\cite{Broedersz:2014}, which were developed to describe reconstituted \emph{in vitro} actin gels with large, $\approx \mu\text{m}$ mesh sizes. {The mechanical response of such networks can largely be understood as the sum of the bending response of individual filament segments comprised between consecutive crosslinkers~\cite{Gardel:2004aa,Storm:2005}. Here we however argue that this framework fails} to account for the elasticity of the much denser branched actin found in the cell. Its high density indeed suppresses classical thermal-fluctuations-induced entropic elasticity, while its branched nature renders even athermal elastic network models problematic. To understand the latter shortcoming, consider a set of consecutive point-like vertices constrained by springs and hinges, which respectively account for the stretching and bending stiffness of a network's filaments. In this network model, a large system acquires a linear elastic modulus only when the number of its spring- and hinge-derived constraints exceeds the number of degrees of freedom of its vertices~\cite{Maxwell:1864,Feng:1985}. This cannot happen in a branched network, as a new branch introduces as many degrees of freedom as constraints [see Fig.~\ref{fig:basics}(A) and demonstration in the SM text]. Due to its low coordination, network models predict it to have a vanishing linear elastic modulus, in apparent contradiction with its measured large stiffness.

\subsection*{Branched network elasticity is atypical}
To clarify the origin of this stiffness, we polymerize branched actin between cylindrical, $\varnothing=6\,\mu\text{m}$ superparamagnetic particles and modulate the external magnetic field to measure the mechanical response of the actin network [Fig.~\ref{fig:basics}(B) and SM]. This microcylinder setup achieves high densities comparable to those found in the cell and obtained in AFM experiments, characterized by mesh sizes $\xi \approx 50\,\text{nm}$~\cite{Bauer:2017}. This approach has a lower force range and a higher throughput than AFM. {This allows us to probe the small-stress response of branched actin with unprecedented resolution and statistics under a wide range of independent experimental conditions}. To avoid introducing residual stresses in the network, we grow branched actin networks to a thickness of several $\mu\text{m}$ under a weak magnetic field, resulting in stresses of the order of a few Pa. Following this polymerization step, we gradually increase the field. This induces an increasing dipolar attraction between the microcylinders and an increasing compression of the network, {but no noticeable widening in the direction transverse to the field}. We assess the network's response by monitoring its deformation as a function of the imposed stress [Fig.~\ref{fig:basics}(C)]. We then compute its tangential elastic modulus as a function of compression. According to elastic network theories, in an overconstrained actin network with similar density this modulus should display a low-stress plateau of the order of $\kappa/\xi^4\approx 6\,\text{kPa}$ at low stress, where $\kappa\simeq 4\times 10^{-26}\,\text{J}\cdot\text{m}$ is the actin bending modulus. {Figure~\ref{fig:basics}(D) shows the contrast between this expectation and our networks' elastic moduli. At low stress, they decrease to values almost two orders of magnitude smaller, down to levels comparable to the detection limit of our setup} It moreover does not display a clear low-stress plateau, confirming that branched actin is too weakly constrained to draw its elasticity from filament stretching and bending alone.

Increasing compressive stresses lead to a rapid increase of the tangential modulus [Fig.~\ref{fig:basics}(D)] unaccounted for by elastic network models. We reason that within our dense networks, even a minute compression causes filaments to collide with one another. The resulting points of contact should then constrain the filaments and enhance the network's mechanical response, as illustrated in Fig.~\ref{fig:basics}(E). To test this interpretation, we conduct numerical simulations of a branched network initially devoid of contacts [Fig.~\ref{fig:basics}(F) and SM]. Its tree-like topology ensures that it is underconstrained. As a point of reference, we first make contacts mechanically irrelevant by simulating ``phantom'' filaments that can interpenetrate without penalty. Consistent with our discussion, this induces an extremely soft elastic response [Fig.~\ref{fig:basics}(G), grey curve]. By contrast, implementing repulsions between the vertices recapitulates the plateau-less elastic response with stiffening exponent of 1 observed in experiments [{slope of the curve in} Fig.~\ref{fig:basics}(H)], thus confirming the crucial role of interfilament contacts.

\subsection*{Branched networks resemble underconstrained filament tangles}
\begin{figure}[t]
\centering
\includegraphics[width=.5\textwidth]{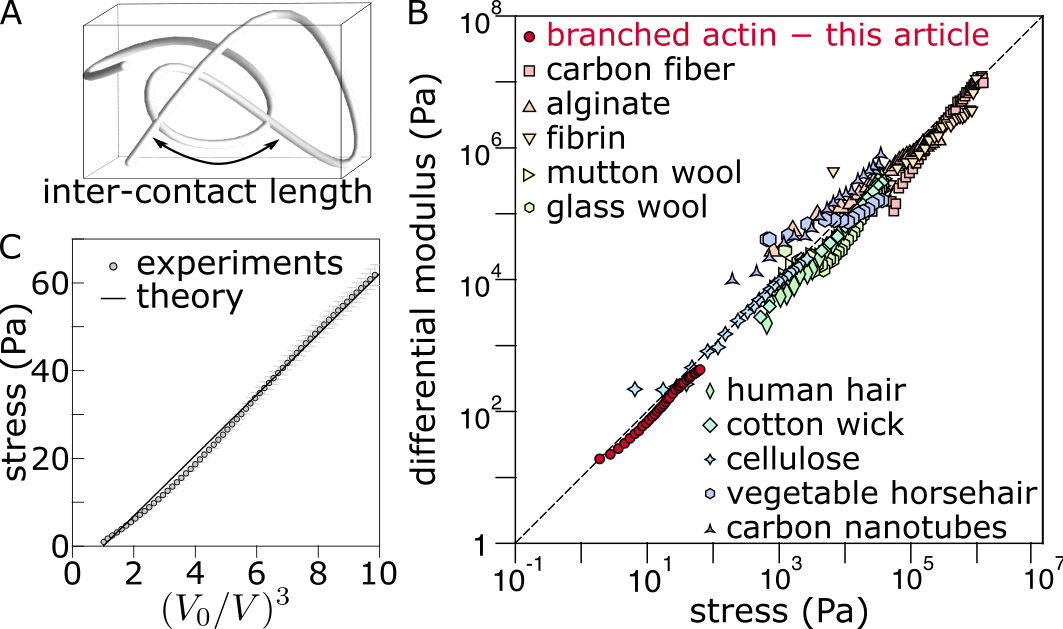}
\caption{\label{fig:nonlinear}
\textbf{Branched network elasticity strongly resembles that of fiber packings, which stiffen due to the proliferation of filament contacts.}
(\textbf{A})~Compression of the packing generically induces such a proliferation, leading to a decrease of the typical inter-contact length.
(\textbf{B})~Similar stiffening behaviors are observed in networks made of fibers whose lengths cover over five orders of magnitude~\cite{Mancini:1999,Poquillon:2005,Masse:2013,Kim:2014,Lopez-Sanchez:2015}.
(\textbf{C})~Equation~\eqref{eq:van~Wyk} is remarkably well verified in our networks {($N = 89$; bars in all figures show standard error; all actin data in this figure is from networks grown in the presence of $500\,\text{nM}$ cofilin).}
}
\end{figure}

While little discussed in cytoskeletal biophysics, contact-induced elasticity in underconstrained systems is a familiar notion in textile mechanics. Consider a collection of long, crumpled textile fibers confined in a volume $V$. As the volume is reduced, the filaments are pushed together and form more contacts Fig.~\ref{fig:nonlinear}(A). As the typical length of the segments comprised between two such contacts decreases, bending them becomes increasingly difficult and the system stiffens.
Contact-driven mechanics additionally leads to a robust stiffening exponent of 1 under compression and the absence of a clear linear elastic plateau in systems without residual stresses [Fig.~\ref{fig:basics} (D) and (H)]. To highlight this point, in Fig.~\ref{fig:nonlinear}(B) we superimpose the stiffening curve of our branched actin with data obtained by uniaxially compressing a wide variety of macroscopic fiber packings {(see SM for mehtodology)}. The resulting parameter-less collapse over more than six orders of magnitude in stress suggests that the analogy with disordered fiber packings is a useful one. At high compression, this mechanical argument results in a simple functional form for the compressive stress $\sigma$ first derived by van~Wyk in 1946 to account for the compressibility of sheep's wool, namely 
\begin{equation}\label{eq:van~Wyk}
\sigma{\propto} (V_0/V)^{3}-1,
\end{equation}
where $V_0$ is the resting volume of the system~\cite{vanWyk:1946}. {Consistent with Fig.~\ref{fig:nonlinear}(B), this relation implies a stiffening exponent of 1 at high compression}.
To {more fully} test the prediction of Eq.~\eqref{eq:van~Wyk}, we grow branched actin in the presence of the protein cofilin. Under the conditions and stoichiometry used in our experiments, cofilin is expected to coat individual actin filaments and decrease their bending rigidity~\cite{McCullough:2008}. These networks are softer than those without cofilin, which allows us to compress them down to smaller volumes and test Eq.~\eqref{eq:van~Wyk} over a broad range of $(V_0/V)^3$. As shown in Fig.~\ref{fig:nonlinear}(C), the affine dependence of Eq.~\eqref{eq:van~Wyk} is remarkably well verified.

\subsection*{Stressed growth locks contacts into the network structure}
\begin{figure}[t!]
\centering
\includegraphics[width=.99\columnwidth]{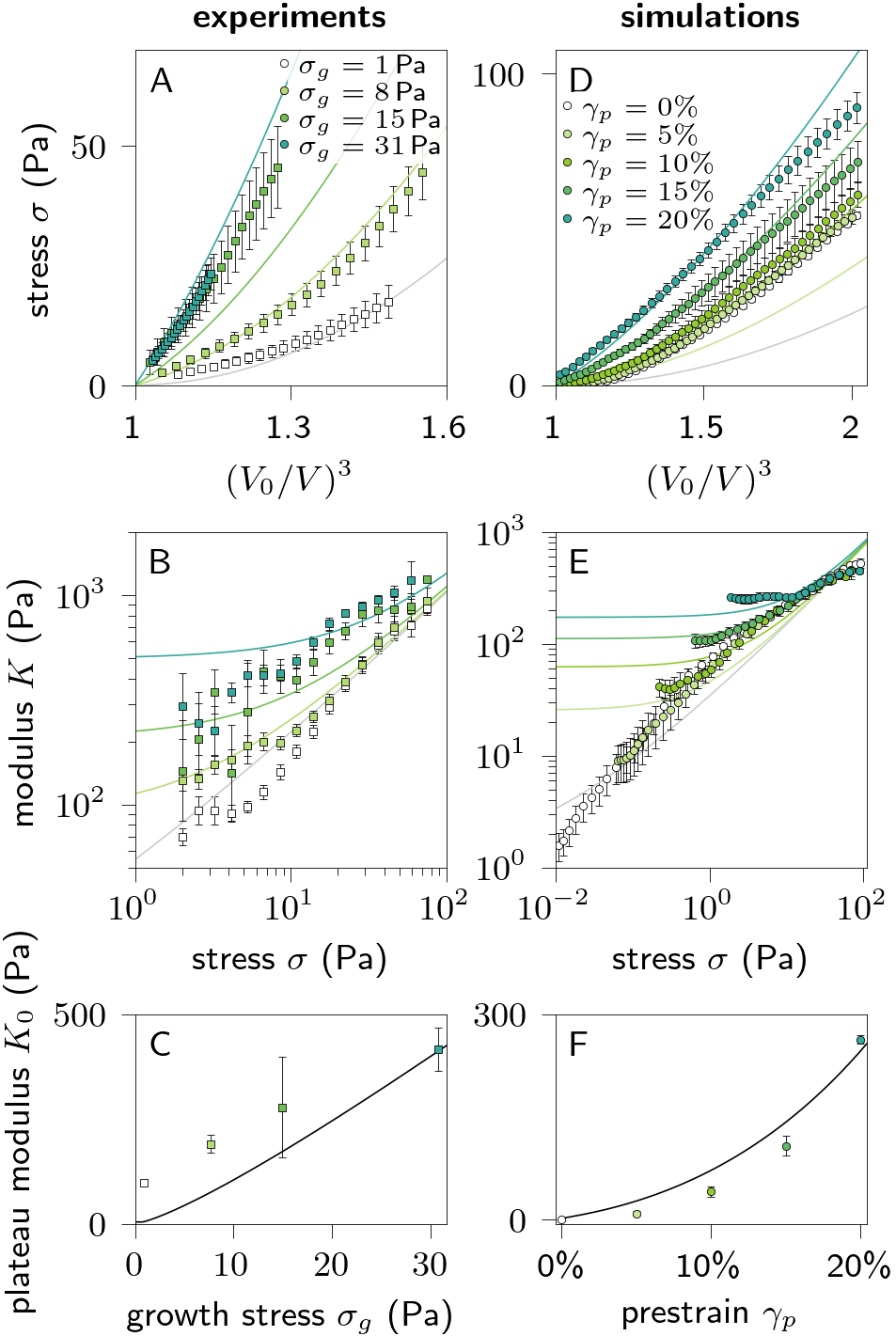}
\caption{\label{fig:stress}
\textbf{Branched networks are stiffer when grown under stress.}
This stiffening captured in our contact-based simulations and analytical model, which uses only three global parameters to collectively fit all experiments, and three for all simulations. \emph{Solid lines} represent the analytical predictions from Eq. \ref{eq:modified_van_Wyk} throughout.s
(\textbf{A})~Experimental compression curves using the representation of Fig.~\ref{fig:nonlinear}(C), although with a smaller horizontal range. The asymptotic $V_0/V\gg 1$ behavior recapitulates the affine prediction of Eq.~\eqref{eq:van~Wyk} ({$N = 46$, $N \geq 6$ per condition).}
(\textbf{B})~Stiffness-stress representation of the same data. Theoretical curves are obtained by simultaneously fitting all $4+4=8$ curves in both panels, yielding parameters $n\simeq 1.02$, ${\kappa}/{\ell^3d}\simeq 606\,\text{Pa}$ and $\Sigma\simeq 17.3\,\text{Pa}$.
(\textbf{C})~Larger growth stresses result in a detectable elastic plateau at low stresses, in agreement with theory.
(\textbf{D})~Compression curves for simulated networks with built-in entanglements added through a phantomized prestrain $\gamma_p$ recapitulate the behavior of their experimental counterparts. ($N=3$ per condition)
(\textbf{E})~A simultaneous fit similar to the one performed in the experiments yields $n=1.02$, ${\kappa}/{\ell^3d}\simeq 195\,\text{Pa}$ and $\alpha_0\simeq 1.11\times |\ln(1-\gamma_p)|$. Here $\alpha_0$ is imposed through a change in volume and not stress, making $\Sigma$ irrelevant.
(\textbf{F})~Consistent with our discussion, build-in entanglements recapitulate the influence of stressed growth on the elastic plateau.
}
\end{figure}

Cells typically assemble branched actin networks in response to opposing stresses exerted at its boundary~\cite{Svitkina:2020}.
To assess the influence of the magnitude $\sigma_g$ of such growth stresses on the network's mechanics, we subject our microcylinders to a controlled magnetic field during actin growth. We thus produce branched networks grown under mean stresses ranging from 0.9 to $30.8\,\text{Pa}$. We next release this growth stress and allow the networks to relax before characterizing their nonlinear elastic response. We observe a dramatic stiffening of the networks grown under large compressive stresses [Fig.~\ref{fig:stress}(A)]. Using another representation that focuses on the small stress response [Fig.~\ref{fig:stress}(B)], we see that branched networks grown under stress also develop a stress-dependent linear elastic plateau. Stressed growth thus restores some aspects of the familiar elastic network models. To explain this behavior, we reason that an opposing stress forces actin filaments to bend during growth. Bent filaments entangle with one another, and form many contacts that subsist after the growth stress is released. To confirm this stiffening mechanism, we introduce a controlled amount of entanglements in our simulations. We thus first turn off interfilament repulsions, then compress the network to cause the resulting phantom filaments to bend and move through each other. Once the interpenetrating filaments are mechanically equilibrated, we turn their interfilament repulsions back on, thus locking the filament entanglements into the network. As we relax the external compression, the networks relax only partially and acquire interfilament contacts which endow them with very similar mechanics as their experimental counterparts, as shown in Figs.~\ref{fig:stress}(D-E).

To capture the effects of this contact accumulation under stressed growth in a simple mean-field model, we extend Eq.~\eqref{eq:van~Wyk} to include two additional effects. First, we replace the infinitely long filaments of the van~Wyk model by finite-length branches. Each such branch may have fewer constraints than degrees of freedom; we denote by $n$ the number of such missing constraints per branch in the absence of interfilament contacts. In weakly compressed networks with few contacts, many branches have too few contacts to be stabilized mechanically. They thus do not participate in the network's mechanics, which results in a soft, initial response that accounts for the small slope at low compression ($V_0/V$ small) in Figs.~\ref{fig:stress}(A) and (D). The network then stiffens more rapidly than expected from Eq.~\eqref{eq:van~Wyk} as an increasing number of branches become stabilized upon increasing compression. Equation~\eqref{eq:van~Wyk} thus becomes
\begin{equation}\label{eq:modified_van_Wyk}
\sigma \approx \frac{\kappa}{\ell^3d}\left[S_n(\alpha_0V_0/V)-S_n(\alpha_0)\right].
\end{equation}
Here the parameter $\alpha_0$ is the number of contacts per branch in the absence of imposed stress (\emph{i.e.}, for $\sigma=0$), $d=8\,\text{nm}$ is the filament diameter and $\ell$ is the mean distance between two branching points along a filament, which is of the order of but potentially distinct from the network mesh size $\xi$. {Similar to the original van Wyk theory, Eq.~\eqref{eq:modified_van_Wyk} is only meant to model the compressive $\sigma\geq 0$ regime.
The full definition of the dimensionless function $S_n(\alpha)$ in Eq.~\eqref{eq:modified_van_Wyk} is detailed in the SM. Qualitatively, it
interpolates between two regimes. In the small-compression regime ($V_0-V\ll V_0$), the network response depends on the extent to which it is undercoordinated, which is characterized by the parameter $n$.
By contrast, the high-compression regime does not depend on $n$ and recapitulates the $\sigma\propto V^{-3}$ asymptotics of Eq.~\eqref{eq:van~Wyk} for $V\rightarrow 0$: $S_n(\alpha)\sim \alpha^3/3$ for $\alpha\gg 1$. 
In the experimentally important $n=1$ case, $S_n(\alpha)$ vanishes for $\alpha\leq 1$. Just above this threshold we have $S_1(\alpha)\sim_{\alpha\rightarrow 1^+} 8(\alpha-1)^3/3$. The SM presents a simple rational function that approximates the way in which $S_1$ smoothly interpolates between these two asymptotic behaviors. 
The existence of a threshold value for $\alpha$ implies that if the number $\alpha_0$ of trapped contacts per branch is smaller than 1, then the network's differential modulus $-\frac{\text{d}\sigma}{\text{d}\ln V}$ vanishes in the linear response regime $V\rightarrow V_0^-$. By contrast, for $\alpha_0>1$ trapped contacts overconstrain the network and induce a finite linear modulus.}

The second ingredient of our model is an analytical prediction for the dependence of $\alpha_0$ on the growth pressure $\sigma_g$ based on the modeling of the bending of growing filaments subjected to an opposing stress. Qualitatively, $\alpha_0$ increases with increasing $\sigma_g$ and crosses over from an $n$-dependent, weak compression regime into a van-Wyk-like large compression regime at a typical growth stress $\Sigma$. We fit our model to our data using fitting parameters $n$, $\ell$ and $\Sigma$ in common for all experimental curves in Fig.~\ref{fig:stress}(A-B). {While the curves in these two panels are extracted the same data, they respectively highlight the networks' large- and small-compression behaviors. Together, they demonstrate that our model satisfactorily describes both regimes.}
We fit the simulations similarly in Fig.~\ref{fig:stress}(D-E), although since there we mimic the growth stress by our aforementioned pre-compression protocol the stress parameter $\Sigma$ is replaced by a deformation parameter. Despite some local deviations, the behavior of the data is \text{satisfactorily} captured by the model in both cases using only three global parameters to respectively fit eight and ten curves. The optimal value of the fitting parameters indicate a branch length $\ell\simeq 202\,\text{nm}$ in the experiments and $\ell\simeq 253\,\text{nm}$ in the simulations, similar to the values observed in lamellipodial branched networks in living cells~\cite{Vinzenz20122775}
Applying a constraint counting argument to our simulations, we expect $n=1$ due to the absence of constraints hindering the rotation of a daughter branch around its mother in our numerical scheme. The fit indeed yields $n\simeq 1.02$. It moreover indicates a very similar value in the experiments. This could indicate that just like in the simulations, the torsional stiffness of actin may only weakly constrain the mechanics of our system, although other defects in the branched architecture of the network could alternatively result in the same behavior.
Finally, we show in the insets of Figs.~\ref{fig:stress}(C) and (F) that our scheme correctly predicts the elastic plateaus of our network without any further fitting parameters.

\begin{figure}[t!]
\centering
\includegraphics[width=.99\columnwidth]{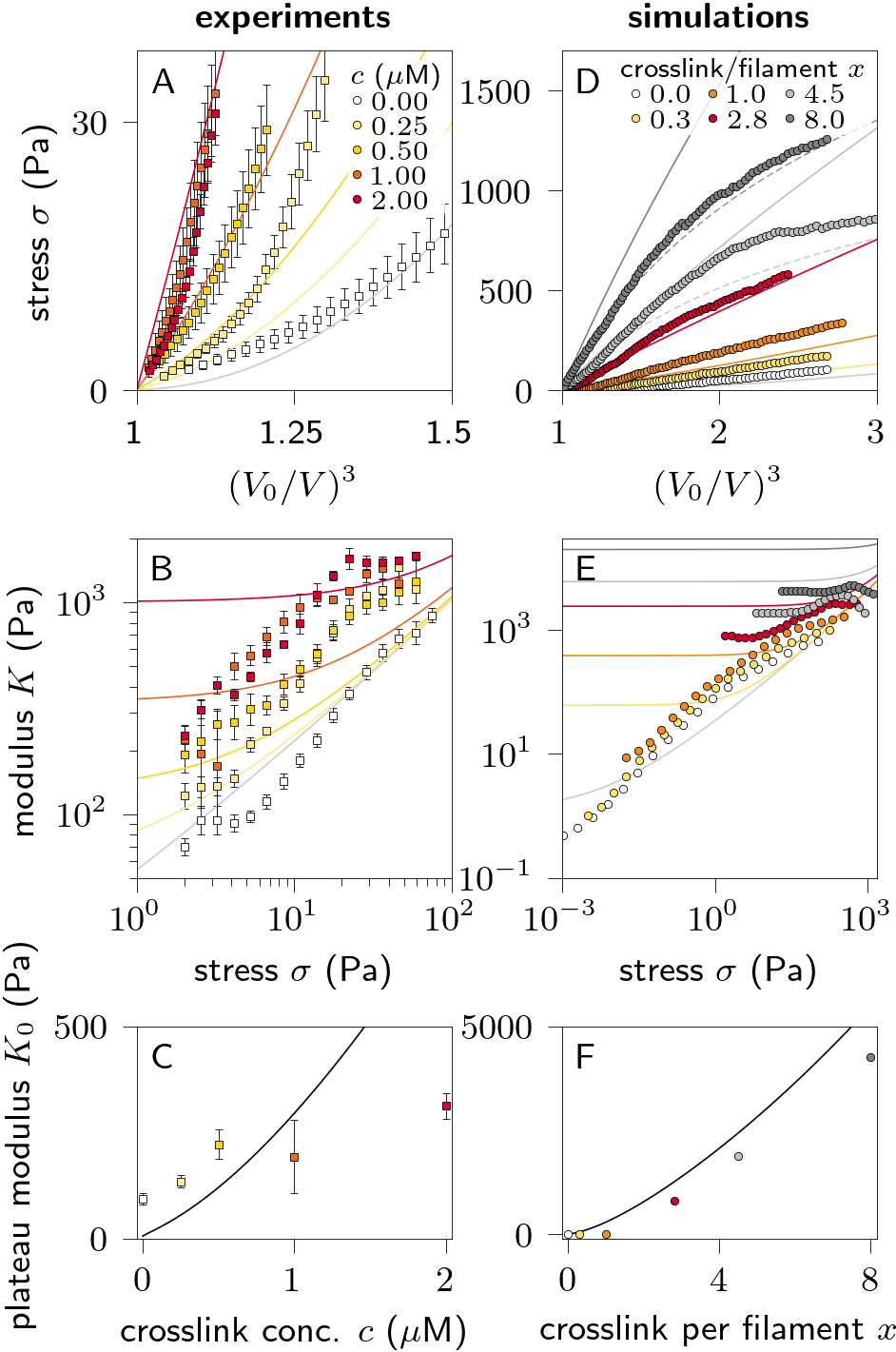}
\caption{\label{fig:crosslink}
\textbf{Crosslinks act as additional contacts, although only at relatively low concentrations.}
The fits here parallel Fig.~\ref{fig:stress} and reuse its parameters, adding only one single global fitting parameter.
(\textbf{A}-\textbf{B})~Increasing the crosslinker concentration $c$ stiffens branched networks. Our only fitting parameter is the number of additional contacts per branch added per unit of $c$, which we find to be equal to $0.16\,\mu\text{M}^{-1}$. Thus networks with $c<6.1\,\mu\text{M}$ should remain underconstrained {($N = 31, N \geq 5$ per condition, all with $\sigma_g\simeq 1\,\text{Pa}$).}
(\textbf{C})~High crosslinker concentrations affect the experiments less than predicted by the model, likely due to a saturation of the incorporation of the crosslinkers into the network.
(\textbf{D}-\textbf{E})~Response of simulated networks with a known number $x$ of crosslinkers per branch. As a result of our fit, we find that each unit of $x$ contributes $0.36$ contacts per branch, and so that networks with $x>2.7$ are overconstrained. This is the case of our two most crosslinked network, while our third most crosslinked is a marginal case. The former two indeed exhibit a response more consistent with a standard linear elastic response (\emph{dashed lines}) than with our underconstrained model (\emph{solid lines}). We thus ignore these two curves in our fit.
(\textbf{F})~A saturation of the incorporation of crosslinkers cannot occur in our simulations, and the resulting increase of modulus with increasing concentration matches the theory much more closely than in the experiments.
}
\end{figure}

\subsection*{In the underconstrained regime, crosslinkers act as obligatory contacts}
Our model implicitly likens trapped contacts to crosslinkers. To test the validity of this analogy, we grow branched networks under low stress in the presence of the cytoskeletal protein alpha-actinin, which crosslinks actin filaments into networks and bundles and is involved in many aspects of cell motility and structure~\cite{Courson:2010}. We investigate crosslinker concentrations ranging from 0 (below the physiological concentration) to $2\,\mu\text{M}$ (well above it).
As shown in Fig.~\ref{fig:crosslink}({A}-B), this results in a stiffening comparable to that resulting from stressed growth. To account for it in our theory, we assume that the addition of crosslinkers is equivalent to the introduction of a number of irreversible contacts proportional to the crosslinker concentration. Reusing the parameters fitted in Fig.~\ref{fig:stress}(A-B), we use the corresponding proportionality constant as our only fitting parameter across 10 curves in Fig.~\ref{fig:crosslink}(A-B) and find a good agreement with the experimental data. Our model however overestimates the magnitude of the crosslinker-induced stiffening at high concentrations [Fig.~\ref{fig:crosslink}(C)], as would be the case if the density of crosslinkers actually incorporated in the experimental network would saturate at high concentration. To probe the effect of better controlled, larger crosslinker densities as well as stronger compression than our experiments allow, we incorporate up to as many as 8 crosslinkers per branch in our simulations and fit our model using the same protocol [Fig.~\ref{fig:crosslink}(D-E)]. As expected, both the low-crosslinker and the low-force response of our networks remains consistent with our contact-induced elasticity theory, thus validating our interpretation that the discrepancies seen in Fig.~\ref{fig:crosslink}(C) were due to insufficient crosslinker incorporation. In addition, we predict and verify that networks where the number of crosslinkers per branch exceeds 2.7 deviate from our model and start resembling the linear elastic response predicted by elastic network theories. 

\subsection*{Number of contacts predicts the stiffness of underconstrained networks}
\begin{figure}[t]
\centering
\includegraphics[width=\columnwidth]{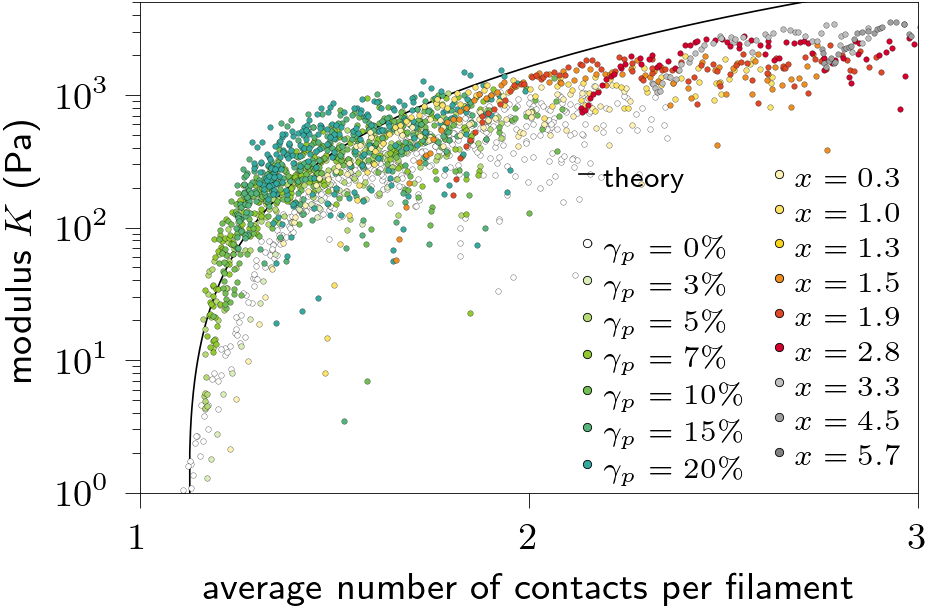}
\caption{\label{fig:contacts}
\textbf{The number of contacts predicts the elastic modulus in all our numerical simulations.}
The initial number of contacts for each system is estimated from the fitting parameters, and the additional contacts are directly counted in the simulations (see SM for details). The relationship predicted by theory is widely valid and breaks down only for heavily crosslinked networks (\emph{gray circles}), as discussed in Fig.~\ref{fig:stress}(G-H). Some prestresses and crosslinking values shown here were omitted in previous figures for the sake of legibility.
}
\end{figure}

As a final challenge to our understanding of branched network elasticity, we note that our theory directly links the elastic modulus of the network to the density of its interfilament contacts, be they induced by a reversible compression of the network, trapped entanglements or crosslinking. By directly counting the number of contacts in all simulations presented above, we indeed find in Fig.~\ref{fig:contacts} that regardless of their origin, the number of contacts predicts the rigidity of the network consistent with theory.

\subsection*{Discussion}

The good agreement of our experimental, numerical and theoretical data under multiple conditions with remarkably few parameters indicate that branched actin elasticity is chiefly based on labile interfilament contacts, and is thus markedly different from that of crosslinked actin. Its undercoordinated nature endows it with mechanical properties that are finely tuned to its growth conditions: growth stresses entangle the filaments and make the network sturdier in much the same way that knitted wool is more rigid than loose wool. This mechanism could endow physiological networks with adaptive characteristics beyond those stemming from a direct regulation by the cell.
According to our measurements, the addition of crosslinkers in physiological concentrations, {\emph{i.e.}, a few hundred nM at most,} hardly affect the mechanics of branched actin. Even the networks obtained at higher concentrations ($\approx 1$ crosslinker per branch) are still well described by contact-induced elasticity. Contact-induced elasticity may be harnessed by the cell to effect rapid and substantial changes in the network's elastic properties through limited actin severing and debranching by relaxing some of the entanglement-induced residual stress. More broadly, our work contrast with many previous \emph{in vitro} physical studies of cytoskeletal networks and the extracellular matrix, which for technical reasons consider networks much more dilute than their physiological counterparts. The high-density regime we consider makes interfilament contacts more frequent and influential, and the ability of our setup to probe low-stress grown networks compared to AFM experiments~\cite{Bieling:2016,Li:2022} allow us to detect their specific mechanical signatures for the first time. While the undercoordinated nature of the branched networks studied here makes them particularly susceptible to this influence, future work may reveal that they constitute a crucial, yet overlooked aspect of cellular mechanics at large.


\begin{thebibliography}{27}%
\makeatletter
\providecommand \@ifxundefined [1]{%
 \@ifx{#1\undefined}
}%
\providecommand \@ifnum [1]{%
 \ifnum #1\expandafter \@firstoftwo
 \else \expandafter \@secondoftwo
 \fi
}%
\providecommand \@ifx [1]{%
 \ifx #1\expandafter \@firstoftwo
 \else \expandafter \@secondoftwo
 \fi
}%
\providecommand \natexlab [1]{#1}%
\providecommand \enquote  [1]{``#1''}%
\providecommand \bibnamefont  [1]{#1}%
\providecommand \bibfnamefont [1]{#1}%
\providecommand \citenamefont [1]{#1}%
\providecommand \href@noop [0]{\@secondoftwo}%
\providecommand \href [0]{\begingroup \@sanitize@url \@href}%
\providecommand \@href[1]{\@@startlink{#1}\@@href}%
\providecommand \@@href[1]{\endgroup#1\@@endlink}%
\providecommand \@sanitize@url [0]{\catcode `\\12\catcode `\$12\catcode
  `\&12\catcode `\#12\catcode `\^12\catcode `\_12\catcode `\%12\relax}%
\providecommand \@@startlink[1]{}%
\providecommand \@@endlink[0]{}%
\providecommand \url  [0]{\begingroup\@sanitize@url \@url }%
\providecommand \@url [1]{\endgroup\@href {#1}{\urlprefix }}%
\providecommand \urlprefix  [0]{URL }%
\providecommand \Eprint [0]{\href }%
\providecommand \doibase [0]{https://doi.org/}%
\providecommand \selectlanguage [0]{\@gobble}%
\providecommand \bibinfo  [0]{\@secondoftwo}%
\providecommand \bibfield  [0]{\@secondoftwo}%
\providecommand \translation [1]{[#1]}%
\providecommand \BibitemOpen [0]{}%
\providecommand \bibitemStop [0]{}%
\providecommand \bibitemNoStop [0]{.\EOS\space}%
\providecommand \EOS [0]{\spacefactor3000\relax}%
\providecommand \BibitemShut  [1]{\csname bibitem#1\endcsname}%
\let\auto@bib@innerbib\@empty
\bibitem [{\citenamefont {Blanchoin}\ \emph {et~al.}(2014)\citenamefont
  {Blanchoin}, \citenamefont {Boujemaa-Paterski}, \citenamefont {Sykes},\ and\
  \citenamefont {Plastino}}]{Blanchoin:2014}%
  \BibitemOpen
  \bibfield  {author} {\bibinfo {author} {\bibfnamefont {L.}~\bibnamefont
  {Blanchoin}}, \bibinfo {author} {\bibfnamefont {R.}~\bibnamefont
  {Boujemaa-Paterski}}, \bibinfo {author} {\bibfnamefont {C.}~\bibnamefont
  {Sykes}},\ and\ \bibinfo {author} {\bibfnamefont {J.}~\bibnamefont
  {Plastino}},\ }\bibfield  {title} {\bibinfo {title} {Actin dynamics,
  architecture, and mechanics in cell motility},\ }\href
  {https://doi.org/10.1152/physrev.00018.2013} {\bibfield  {journal} {\bibinfo
  {journal} {Physiol. Rev.}\ }\textbf {\bibinfo {volume} {94}},\ \bibinfo
  {pages} {235} (\bibinfo {year} {2014})}\BibitemShut {NoStop}%
\bibitem [{\citenamefont {Marcy}\ \emph {et~al.}(2004)\citenamefont {Marcy},
  \citenamefont {Prost}, \citenamefont {Carlier},\ and\ \citenamefont
  {Sykes}}]{Marcy:2004}%
  \BibitemOpen
  \bibfield  {author} {\bibinfo {author} {\bibfnamefont {Y.}~\bibnamefont
  {Marcy}}, \bibinfo {author} {\bibfnamefont {J.}~\bibnamefont {Prost}},
  \bibinfo {author} {\bibfnamefont {M.-F.}\ \bibnamefont {Carlier}},\ and\
  \bibinfo {author} {\bibfnamefont {C.}~\bibnamefont {Sykes}},\ }\bibfield
  {title} {\bibinfo {title} {Forces generated during actin-based propulsion: A
  direct measurement by micromanipulation},\ }\href
  {https://doi.org/10.1073/pnas.0307704101} {\bibfield  {journal} {\bibinfo
  {journal} {Proc. Natl. Acad. Sci. U.S.A.}\ }\textbf {\bibinfo {volume}
  {101}},\ \bibinfo {pages} {5992} (\bibinfo {year} {2004})}\BibitemShut
  {NoStop}%
\bibitem [{\citenamefont {Chaudhuri}\ \emph {et~al.}(2007)\citenamefont
  {Chaudhuri}, \citenamefont {Parekh},\ and\ \citenamefont
  {Fletcher}}]{Chaudhuri:2007}%
  \BibitemOpen
  \bibfield  {author} {\bibinfo {author} {\bibfnamefont {O.}~\bibnamefont
  {Chaudhuri}}, \bibinfo {author} {\bibfnamefont {S.~H.}\ \bibnamefont
  {Parekh}},\ and\ \bibinfo {author} {\bibfnamefont {D.~A.}\ \bibnamefont
  {Fletcher}},\ }\bibfield  {title} {\bibinfo {title} {Reversible stress
  softening of actin networks},\ }\href {https://doi.org/10.1038/nature05459}
  {\bibfield  {journal} {\bibinfo  {journal} {Nature}\ }\textbf {\bibinfo
  {volume} {445}},\ \bibinfo {pages} {295} (\bibinfo {year}
  {2007})}\BibitemShut {NoStop}%
\bibitem [{\citenamefont {Pujol}\ \emph {et~al.}(2012)\citenamefont {Pujol},
  \citenamefont {du~Roure}, \citenamefont {Fermigier},\ and\ \citenamefont
  {Heuvingh}}]{Pujol:2012}%
  \BibitemOpen
  \bibfield  {author} {\bibinfo {author} {\bibfnamefont {T.}~\bibnamefont
  {Pujol}}, \bibinfo {author} {\bibfnamefont {O.}~\bibnamefont {du~Roure}},
  \bibinfo {author} {\bibfnamefont {M.}~\bibnamefont {Fermigier}},\ and\
  \bibinfo {author} {\bibfnamefont {J.}~\bibnamefont {Heuvingh}},\ }\bibfield
  {title} {\bibinfo {title} {Impact of branching on the elasticity of actin
  networks},\ }\href {https://doi.org/10.1073/pnas.1121238109} {\bibfield
  {journal} {\bibinfo  {journal} {Proc. Natl. Acad. Sci. U.S.A.}\ }\textbf
  {\bibinfo {volume} {109}},\ \bibinfo {pages} {10364} (\bibinfo {year}
  {2012})}\BibitemShut {NoStop}%
\bibitem [{\citenamefont {Bau{\"e}r}\ \emph {et~al.}(2017)\citenamefont
  {Bau{\"e}r}, \citenamefont {Tavacoli}, \citenamefont {Pujol}, \citenamefont
  {Planade}, \citenamefont {Heuvingh},\ and\ \citenamefont
  {du~Roure}}]{Bauer:2017}%
  \BibitemOpen
  \bibfield  {author} {\bibinfo {author} {\bibfnamefont {P.}~\bibnamefont
  {Bau{\"e}r}}, \bibinfo {author} {\bibfnamefont {J.}~\bibnamefont {Tavacoli}},
  \bibinfo {author} {\bibfnamefont {T.}~\bibnamefont {Pujol}}, \bibinfo
  {author} {\bibfnamefont {J.}~\bibnamefont {Planade}}, \bibinfo {author}
  {\bibfnamefont {J.}~\bibnamefont {Heuvingh}},\ and\ \bibinfo {author}
  {\bibfnamefont {O.}~\bibnamefont {du~Roure}},\ }\bibfield  {title} {\bibinfo
  {title} {A new method to measure mechanics and dynamic assembly of branched
  actin networks},\ }\href {https://doi.org/10.1038/s41598-017-15638-5}
  {\bibfield  {journal} {\bibinfo  {journal} {Sci. Rep.}\ }\textbf {\bibinfo
  {volume} {7}},\ \bibinfo {pages} {15688} (\bibinfo {year}
  {2017})}\BibitemShut {NoStop}%
\bibitem [{\citenamefont {Bieling}\ \emph {et~al.}(2016)\citenamefont
  {Bieling}, \citenamefont {Li}, \citenamefont {Weichsel}, \citenamefont
  {McGorty}, \citenamefont {Jreij}, \citenamefont {Huang}, \citenamefont
  {Fletcher},\ and\ \citenamefont {Mullins}}]{Bieling:2016}%
  \BibitemOpen
  \bibfield  {author} {\bibinfo {author} {\bibfnamefont {P.}~\bibnamefont
  {Bieling}}, \bibinfo {author} {\bibfnamefont {T.-D.}\ \bibnamefont {Li}},
  \bibinfo {author} {\bibfnamefont {J.}~\bibnamefont {Weichsel}}, \bibinfo
  {author} {\bibfnamefont {R.}~\bibnamefont {McGorty}}, \bibinfo {author}
  {\bibfnamefont {P.}~\bibnamefont {Jreij}}, \bibinfo {author} {\bibfnamefont
  {B.}~\bibnamefont {Huang}}, \bibinfo {author} {\bibfnamefont {D.~A.}\
  \bibnamefont {Fletcher}},\ and\ \bibinfo {author} {\bibfnamefont {R.~D.}\
  \bibnamefont {Mullins}},\ }\bibfield  {title} {\bibinfo {title} {Force
  feedback controls motor activity and mechanical properties of self-assembling
  branched actin networks},\ }\href
  {https://doi.org/10.1016/j.cell.2015.11.057} {\bibfield  {journal} {\bibinfo
  {journal} {Cell}\ }\textbf {\bibinfo {volume} {164}},\ \bibinfo {pages} {115}
  (\bibinfo {year} {2016})}\BibitemShut {NoStop}%
\bibitem [{\citenamefont {Mueller}\ \emph {et~al.}(2017)\citenamefont
  {Mueller}, \citenamefont {Szep}, \citenamefont {Nemethova}, \citenamefont
  {de~Vries}, \citenamefont {Lieber}, \citenamefont {Winkler}, \citenamefont
  {Kruse}, \citenamefont {Small}, \citenamefont {Schmeiser}, \citenamefont
  {Keren}, \citenamefont {Hauschild},\ and\ \citenamefont
  {Sixt}}]{Mueller:2017}%
  \BibitemOpen
  \bibfield  {author} {\bibinfo {author} {\bibfnamefont {J.}~\bibnamefont
  {Mueller}}, \bibinfo {author} {\bibfnamefont {G.}~\bibnamefont {Szep}},
  \bibinfo {author} {\bibfnamefont {M.}~\bibnamefont {Nemethova}}, \bibinfo
  {author} {\bibfnamefont {I.}~\bibnamefont {de~Vries}}, \bibinfo {author}
  {\bibfnamefont {A.~D.}\ \bibnamefont {Lieber}}, \bibinfo {author}
  {\bibfnamefont {C.}~\bibnamefont {Winkler}}, \bibinfo {author} {\bibfnamefont
  {K.}~\bibnamefont {Kruse}}, \bibinfo {author} {\bibfnamefont {J.~V.}\
  \bibnamefont {Small}}, \bibinfo {author} {\bibfnamefont {C.}~\bibnamefont
  {Schmeiser}}, \bibinfo {author} {\bibfnamefont {K.}~\bibnamefont {Keren}},
  \bibinfo {author} {\bibfnamefont {R.}~\bibnamefont {Hauschild}},\ and\
  \bibinfo {author} {\bibfnamefont {M.}~\bibnamefont {Sixt}},\ }\bibfield
  {title} {\bibinfo {title} {Load adaptation of lamellipodial actin networks},\
  }\href {https://doi.org/10.1016/j.cell.2017.07.051} {\bibfield  {journal}
  {\bibinfo  {journal} {Cell}\ }\textbf {\bibinfo {volume} {171}},\ \bibinfo
  {pages} {188} (\bibinfo {year} {2017})}\BibitemShut {NoStop}%
\bibitem [{\citenamefont {Mullins}\ \emph {et~al.}(1998)\citenamefont
  {Mullins}, \citenamefont {Heuser},\ and\ \citenamefont
  {Pollard}}]{Mullins98}%
  \BibitemOpen
  \bibfield  {author} {\bibinfo {author} {\bibfnamefont {R.~D.}\ \bibnamefont
  {Mullins}}, \bibinfo {author} {\bibfnamefont {J.~A.}\ \bibnamefont
  {Heuser}},\ and\ \bibinfo {author} {\bibfnamefont {T.~D.}\ \bibnamefont
  {Pollard}},\ }\bibfield  {title} {\bibinfo {title} {The interaction of arp2/3
  complex with actin: Nucleation, high affinity pointed end capping, and
  formation of branching networks of filaments},\ }\href
  {https://doi.org/10.1073/pnas.95.11.6181} {\bibfield  {journal} {\bibinfo
  {journal} {Proceedings of the National Academy of Sciences}\ }\textbf
  {\bibinfo {volume} {95}},\ \bibinfo {pages} {6181} (\bibinfo {year}
  {1998})},\ \Eprint
  {https://arxiv.org/abs/https://www.pnas.org/doi/pdf/10.1073/pnas.95.11.6181}
  {https://www.pnas.org/doi/pdf/10.1073/pnas.95.11.6181} \BibitemShut {NoStop}%
\bibitem [{\citenamefont {Loisel}\ \emph {et~al.}(1999)\citenamefont {Loisel},
  \citenamefont {Boujemaa-Paterski}, \citenamefont {Pantaloni},\ and\
  \citenamefont {Carlier}}]{Loisel1999}%
  \BibitemOpen
  \bibfield  {author} {\bibinfo {author} {\bibfnamefont {T.~P.}\ \bibnamefont
  {Loisel}}, \bibinfo {author} {\bibfnamefont {R.}~\bibnamefont
  {Boujemaa-Paterski}}, \bibinfo {author} {\bibfnamefont {D.}~\bibnamefont
  {Pantaloni}},\ and\ \bibinfo {author} {\bibfnamefont {M.-F.}\ \bibnamefont
  {Carlier}},\ }\bibfield  {title} {\bibinfo {title} {{Reconstitution of
  actin-based motility of Listeria and Shigella using pure proteins.}},\ }\href
  {https://doi.org/10.1038/44183} {\bibfield  {journal} {\bibinfo  {journal}
  {Nature}\ }\textbf {\bibinfo {volume} {401}},\ \bibinfo {pages} {613}
  (\bibinfo {year} {1999})}\BibitemShut {NoStop}%
\bibitem [{\citenamefont {Lappalainen}\ \emph {et~al.}(2022)\citenamefont
  {Lappalainen}, \citenamefont {Kotila}, \citenamefont {J{\'e}gou},\ and\
  \citenamefont {Romet-Lemonne}}]{Lappalainen2022}%
  \BibitemOpen
  \bibfield  {author} {\bibinfo {author} {\bibfnamefont {P.}~\bibnamefont
  {Lappalainen}}, \bibinfo {author} {\bibfnamefont {T.}~\bibnamefont {Kotila}},
  \bibinfo {author} {\bibfnamefont {A.}~\bibnamefont {J{\'e}gou}},\ and\
  \bibinfo {author} {\bibfnamefont {G.}~\bibnamefont {Romet-Lemonne}},\
  }\bibfield  {title} {\bibinfo {title} {Biochemical and mechanical regulation
  of actin dynamics},\ }\href@noop {} {\bibfield  {journal} {\bibinfo
  {journal} {Nature Reviews Molecular Cell Biology}\ }\textbf {\bibinfo
  {volume} {23}},\ \bibinfo {pages} {836} (\bibinfo {year} {2022})}\BibitemShut
  {NoStop}%
\bibitem [{\citenamefont {Risca}\ \emph {et~al.}(2012)\citenamefont {Risca},
  \citenamefont {Wang}, \citenamefont {Chaudhuri}, \citenamefont {Chia},
  \citenamefont {Geissler},\ and\ \citenamefont {Fletcher}}]{Risca:2012}%
  \BibitemOpen
  \bibfield  {author} {\bibinfo {author} {\bibfnamefont {V.~I.}\ \bibnamefont
  {Risca}}, \bibinfo {author} {\bibfnamefont {E.~B.}\ \bibnamefont {Wang}},
  \bibinfo {author} {\bibfnamefont {O.}~\bibnamefont {Chaudhuri}}, \bibinfo
  {author} {\bibfnamefont {J.~J.}\ \bibnamefont {Chia}}, \bibinfo {author}
  {\bibfnamefont {P.~L.}\ \bibnamefont {Geissler}},\ and\ \bibinfo {author}
  {\bibfnamefont {D.~A.}\ \bibnamefont {Fletcher}},\ }\bibfield  {title}
  {\bibinfo {title} {Actin filament curvature biases branching direction},\
  }\href {https://doi.org/10.1073/pnas.1114292109} {\bibfield  {journal}
  {\bibinfo  {journal} {Proc. Natl. Acad. Sci. U.S.A.}\ }\textbf {\bibinfo
  {volume} {109}},\ \bibinfo {pages} {2913} (\bibinfo {year}
  {2012})}\BibitemShut {NoStop}%
\bibitem [{\citenamefont {Li}\ \emph {et~al.}(2022)\citenamefont {Li},
  \citenamefont {Bieling}, \citenamefont {Weichsel}, \citenamefont {Mullins},\
  and\ \citenamefont {Fletcher}}]{Li:2022}%
  \BibitemOpen
  \bibfield  {author} {\bibinfo {author} {\bibfnamefont {T.-D.}\ \bibnamefont
  {Li}}, \bibinfo {author} {\bibfnamefont {P.}~\bibnamefont {Bieling}},
  \bibinfo {author} {\bibfnamefont {J.}~\bibnamefont {Weichsel}}, \bibinfo
  {author} {\bibfnamefont {R.~D.}\ \bibnamefont {Mullins}},\ and\ \bibinfo
  {author} {\bibfnamefont {D.~A.}\ \bibnamefont {Fletcher}},\ }\bibfield
  {title} {\bibinfo {title} {The molecular mechanism of load adaptation by
  branched actin networks},\ }\href {https://doi.org/10.7554/eLife.73145}
  {\bibfield  {journal} {\bibinfo  {journal} {{eLife}}\ }\textbf {\bibinfo
  {volume} {11}},\ \bibinfo {pages} {e73145} (\bibinfo {year}
  {2022})}\BibitemShut {NoStop}%
\bibitem [{\citenamefont {Broedersz}\ and\ \citenamefont
  {MacKintosh}(2014)}]{Broedersz:2014}%
  \BibitemOpen
  \bibfield  {author} {\bibinfo {author} {\bibfnamefont {C.~P.}\ \bibnamefont
  {Broedersz}}\ and\ \bibinfo {author} {\bibfnamefont {F.~C.}\ \bibnamefont
  {MacKintosh}},\ }\bibfield  {title} {\bibinfo {title} {Modeling semiflexible
  polymer networks},\ }\href {https://doi.org/10.1103/RevModPhys.86.995}
  {\bibfield  {journal} {\bibinfo  {journal} {Rev. Mod. Phys.}\ }\textbf
  {\bibinfo {volume} {86}},\ \bibinfo {pages} {995} (\bibinfo {year}
  {2014})}\BibitemShut {NoStop}%
\bibitem [{\citenamefont {Gardel}\ \emph {et~al.}(2004)\citenamefont {Gardel},
  \citenamefont {Shin}, \citenamefont {MacKintosh}, \citenamefont {Mahadevan},
  \citenamefont {Matsudaira},\ and\ \citenamefont {Weitz}}]{Gardel:2004aa}%
  \BibitemOpen
  \bibfield  {author} {\bibinfo {author} {\bibfnamefont {M.~L.}\ \bibnamefont
  {Gardel}}, \bibinfo {author} {\bibfnamefont {J.~H.}\ \bibnamefont {Shin}},
  \bibinfo {author} {\bibfnamefont {F.~C.}\ \bibnamefont {MacKintosh}},
  \bibinfo {author} {\bibfnamefont {L.}~\bibnamefont {Mahadevan}}, \bibinfo
  {author} {\bibfnamefont {P.}~\bibnamefont {Matsudaira}},\ and\ \bibinfo
  {author} {\bibfnamefont {D.~A.}\ \bibnamefont {Weitz}},\ }\bibfield  {title}
  {\bibinfo {title} {Elastic behavior of cross-linked and bundled actin
  networks.},\ }\href {https://doi.org/10.1126/science.1095087} {\bibfield
  {journal} {\bibinfo  {journal} {Science}\ }\textbf {\bibinfo {volume}
  {304}},\ \bibinfo {pages} {1301} (\bibinfo {year} {2004})}\BibitemShut
  {NoStop}%
\bibitem [{\citenamefont {Storm}\ \emph {et~al.}(2005)\citenamefont {Storm},
  \citenamefont {Pastore}, \citenamefont {MacKintosh}, \citenamefont
  {Lubensky},\ and\ \citenamefont {Janmey}}]{Storm:2005}%
  \BibitemOpen
  \bibfield  {author} {\bibinfo {author} {\bibfnamefont {C.}~\bibnamefont
  {Storm}}, \bibinfo {author} {\bibfnamefont {J.~J.}\ \bibnamefont {Pastore}},
  \bibinfo {author} {\bibfnamefont {F.~C.}\ \bibnamefont {MacKintosh}},
  \bibinfo {author} {\bibfnamefont {T.~C.}\ \bibnamefont {Lubensky}},\ and\
  \bibinfo {author} {\bibfnamefont {P.~A.}\ \bibnamefont {Janmey}},\ }\bibfield
   {title} {\bibinfo {title} {Nonlinear elasticity in biological gels},\ }\href
  {https://doi.org/10.1038/nature03521} {\bibfield  {journal} {\bibinfo
  {journal} {Nature}\ }\textbf {\bibinfo {volume} {435}},\ \bibinfo {pages}
  {191} (\bibinfo {year} {2005})}\BibitemShut {NoStop}%
\bibitem [{\citenamefont {Maxwell}(1864)}]{Maxwell:1864}%
  \BibitemOpen
  \bibfield  {author} {\bibinfo {author} {\bibfnamefont {J.~C.}\ \bibnamefont
  {Maxwell}},\ }\bibfield  {title} {\bibinfo {title} {On the calculation of the
  equilibrium and stiffness of frames},\ }\href@noop {} {\bibfield  {journal}
  {\bibinfo  {journal} {Philos. Mag.}\ }\textbf {\bibinfo {volume} {27}},\
  \bibinfo {pages} {294} (\bibinfo {year} {1864})}\BibitemShut {NoStop}%
\bibitem [{\citenamefont {Feng}\ \emph {et~al.}(1985)\citenamefont {Feng},
  \citenamefont {Thorpe},\ and\ \citenamefont {Garboczi}}]{Feng:1985}%
  \BibitemOpen
  \bibfield  {author} {\bibinfo {author} {\bibfnamefont {S.}~\bibnamefont
  {Feng}}, \bibinfo {author} {\bibfnamefont {M.~F.}\ \bibnamefont {Thorpe}},\
  and\ \bibinfo {author} {\bibfnamefont {E.}~\bibnamefont {Garboczi}},\
  }\bibfield  {title} {\bibinfo {title} {Effective-medium theory of percolation
  on central-force elastic networks},\ }\href
  {https://doi.org/10.1103/PhysRevB.31.276} {\bibfield  {journal} {\bibinfo
  {journal} {Phys. Rev. B}\ }\textbf {\bibinfo {volume} {31}},\ \bibinfo
  {pages} {276} (\bibinfo {year} {1985})}\BibitemShut {NoStop}%
\bibitem [{\citenamefont {van Wyk}(1946)}]{vanWyk:1946}%
  \BibitemOpen
  \bibfield  {author} {\bibinfo {author} {\bibfnamefont {C.~M.}\ \bibnamefont
  {van Wyk}},\ }\bibfield  {title} {\bibinfo {title} {20---note on the
  compressibility of wool},\ }\href {https://doi.org/10.1080/19447024608659279}
  {\bibfield  {journal} {\bibinfo  {journal} {Journal of the Textile Institute
  Transactions}\ }\textbf {\bibinfo {volume} {37}},\ \bibinfo {pages} {T285}
  (\bibinfo {year} {1946})}\BibitemShut {NoStop}%
\bibitem [{\citenamefont {McCullough}\ \emph {et~al.}(2008)\citenamefont
  {McCullough}, \citenamefont {Blanchoin}, \citenamefont {Martiel},\ and\
  \citenamefont {De~la Cruz}}]{McCullough:2008}%
  \BibitemOpen
  \bibfield  {author} {\bibinfo {author} {\bibfnamefont {B.~R.}\ \bibnamefont
  {McCullough}}, \bibinfo {author} {\bibfnamefont {L.}~\bibnamefont
  {Blanchoin}}, \bibinfo {author} {\bibfnamefont {J.-L.}\ \bibnamefont
  {Martiel}},\ and\ \bibinfo {author} {\bibfnamefont {E.~M.}\ \bibnamefont
  {De~la Cruz}},\ }\bibfield  {title} {\bibinfo {title} {Cofilin increases the
  bending flexibility of actin filaments: implications for severing and cell
  mechanics},\ }\href {https://doi.org/10.1016/j.jmb.2008.05.055} {\bibfield
  {journal} {\bibinfo  {journal} {J. Mol. Biol.}\ }\textbf {\bibinfo {volume}
  {381}},\ \bibinfo {pages} {550} (\bibinfo {year} {2008})}\BibitemShut
  {NoStop}%
\bibitem [{\citenamefont {Mancini}\ \emph {et~al.}(1999)\citenamefont
  {Mancini}, \citenamefont {Moresi},\ and\ \citenamefont
  {Rancini}}]{Mancini:1999}%
  \BibitemOpen
  \bibfield  {author} {\bibinfo {author} {\bibfnamefont {M.}~\bibnamefont
  {Mancini}}, \bibinfo {author} {\bibfnamefont {M.}~\bibnamefont {Moresi}},\
  and\ \bibinfo {author} {\bibfnamefont {R.}~\bibnamefont {Rancini}},\
  }\bibfield  {title} {\bibinfo {title} {Uniaxial compression and stress
  relaxation tests on alginate gels},\ }\href
  {https://doi.org/10.1111/j.1745-4603.1999.tb00235.x} {\bibfield  {journal}
  {\bibinfo  {journal} {Journal of Texture Studies}\ }\textbf {\bibinfo
  {volume} {30}},\ \bibinfo {pages} {639} (\bibinfo {year} {1999})}\BibitemShut
  {NoStop}%
\bibitem [{\citenamefont {Poquillon}\ \emph {et~al.}(2005)\citenamefont
  {Poquillon}, \citenamefont {Viguier},\ and\ \citenamefont
  {Andrieu}}]{Poquillon:2005}%
  \BibitemOpen
  \bibfield  {author} {\bibinfo {author} {\bibfnamefont {D.}~\bibnamefont
  {Poquillon}}, \bibinfo {author} {\bibfnamefont {B.}~\bibnamefont {Viguier}},\
  and\ \bibinfo {author} {\bibfnamefont {E.}~\bibnamefont {Andrieu}},\
  }\bibfield  {title} {\bibinfo {title} {Experimental data about mechanical
  behaviour during compression tests for various matted fibres},\ }\href@noop
  {} {\bibfield  {journal} {\bibinfo  {journal} {Journal of Materials Science}\
  }\textbf {\bibinfo {volume} {40}},\ \bibinfo {pages} {5963} (\bibinfo {year}
  {2005})}\BibitemShut {NoStop}%
\bibitem [{\citenamefont {Masse}\ and\ \citenamefont
  {Poquillon}(2013)}]{Masse:2013}%
  \BibitemOpen
  \bibfield  {author} {\bibinfo {author} {\bibfnamefont {J.-P.}\ \bibnamefont
  {Masse}}\ and\ \bibinfo {author} {\bibfnamefont {D.}~\bibnamefont
  {Poquillon}},\ }\bibfield  {title} {\bibinfo {title} {Mechanical behavior of
  entangled materials with or without cross-linked fibers},\ }\href
  {https://doi.org/10.1016/j.scriptamat.2012.05.047} {\bibfield  {journal}
  {\bibinfo  {journal} {Scripta Materialia}\ }\textbf {\bibinfo {volume}
  {68}},\ \bibinfo {pages} {39} (\bibinfo {year} {2013})}\BibitemShut {NoStop}%
\bibitem [{\citenamefont {Kim}\ \emph {et~al.}(2014)\citenamefont {Kim},
  \citenamefont {Litvinov}, \citenamefont {Weisel},\ and\ \citenamefont
  {Alber}}]{Kim:2014}%
  \BibitemOpen
  \bibfield  {author} {\bibinfo {author} {\bibfnamefont {O.~V.}\ \bibnamefont
  {Kim}}, \bibinfo {author} {\bibfnamefont {R.~I.}\ \bibnamefont {Litvinov}},
  \bibinfo {author} {\bibfnamefont {J.~W.}\ \bibnamefont {Weisel}},\ and\
  \bibinfo {author} {\bibfnamefont {M.~S.}\ \bibnamefont {Alber}},\ }\bibfield
  {title} {\bibinfo {title} {Structural basis for the nonlinear mechanics of
  fibrin networks under compression},\ }\href
  {https://doi.org/10.1016/j.biomaterials.2014.04.056} {\bibfield  {journal}
  {\bibinfo  {journal} {Biomaterials}\ }\textbf {\bibinfo {volume} {35}},\
  \bibinfo {pages} {6739} (\bibinfo {year} {2014})}\BibitemShut {NoStop}%
\bibitem [{\citenamefont {Lopez-Sanchez}\ \emph {et~al.}(2015)\citenamefont
  {Lopez-Sanchez}, \citenamefont {Cersosimo}, \citenamefont {Wang},
  \citenamefont {Flanagan}, \citenamefont {Stokes},\ and\ \citenamefont
  {Gidley}}]{Lopez-Sanchez:2015}%
  \BibitemOpen
  \bibfield  {author} {\bibinfo {author} {\bibfnamefont {P.}~\bibnamefont
  {Lopez-Sanchez}}, \bibinfo {author} {\bibfnamefont {J.}~\bibnamefont
  {Cersosimo}}, \bibinfo {author} {\bibfnamefont {D.}~\bibnamefont {Wang}},
  \bibinfo {author} {\bibfnamefont {B.}~\bibnamefont {Flanagan}}, \bibinfo
  {author} {\bibfnamefont {J.~R.}\ \bibnamefont {Stokes}},\ and\ \bibinfo
  {author} {\bibfnamefont {M.~J.}\ \bibnamefont {Gidley}},\ }\bibfield  {title}
  {\bibinfo {title} {Poroelastic mechanical effects of hemicelluloses on
  cellulosic hydrogels under compression},\ }\href
  {https://doi.org/10.1371/journal.pone.0122132} {\bibfield  {journal}
  {\bibinfo  {journal} {PLoS One}\ }\textbf {\bibinfo {volume} {10}},\ \bibinfo
  {pages} {e0122132} (\bibinfo {year} {2015})}\BibitemShut {NoStop}%
\bibitem [{\citenamefont {Svitkina}(2020)}]{Svitkina:2020}%
  \BibitemOpen
  \bibfield  {author} {\bibinfo {author} {\bibfnamefont {T.~M.}\ \bibnamefont
  {Svitkina}},\ }\bibfield  {title} {\bibinfo {title} {Actin cell cortex:
  Structure and molecular organization},\ }\href
  {https://doi.org/10.1016/j.tcb.2020.03.005} {\bibfield  {journal} {\bibinfo
  {journal} {Trends Cell Biol}\ }\textbf {\bibinfo {volume} {30}},\ \bibinfo
  {pages} {556} (\bibinfo {year} {2020})}\BibitemShut {NoStop}%
\bibitem [{\citenamefont {Vinzenz}\ \emph {et~al.}(2012)\citenamefont
  {Vinzenz}, \citenamefont {Nemethova}, \citenamefont {Schur}, \citenamefont
  {Mueller}, \citenamefont {Narita}, \citenamefont {Urban}, \citenamefont
  {Winkler}, \citenamefont {Schmeiser}, \citenamefont {Koestler}, \citenamefont
  {Rottner}, \citenamefont {Resch}, \citenamefont {Maeda},\ and\ \citenamefont
  {Small}}]{Vinzenz20122775}%
  \BibitemOpen
  \bibfield  {author} {\bibinfo {author} {\bibfnamefont {M.}~\bibnamefont
  {Vinzenz}}, \bibinfo {author} {\bibfnamefont {M.}~\bibnamefont {Nemethova}},
  \bibinfo {author} {\bibfnamefont {F.}~\bibnamefont {Schur}}, \bibinfo
  {author} {\bibfnamefont {J.}~\bibnamefont {Mueller}}, \bibinfo {author}
  {\bibfnamefont {A.}~\bibnamefont {Narita}}, \bibinfo {author} {\bibfnamefont
  {E.}~\bibnamefont {Urban}}, \bibinfo {author} {\bibfnamefont
  {C.}~\bibnamefont {Winkler}}, \bibinfo {author} {\bibfnamefont
  {C.}~\bibnamefont {Schmeiser}}, \bibinfo {author} {\bibfnamefont {S.~A.}\
  \bibnamefont {Koestler}}, \bibinfo {author} {\bibfnamefont {K.}~\bibnamefont
  {Rottner}}, \bibinfo {author} {\bibfnamefont {G.~P.}\ \bibnamefont {Resch}},
  \bibinfo {author} {\bibfnamefont {Y.}~\bibnamefont {Maeda}},\ and\ \bibinfo
  {author} {\bibfnamefont {J.~V.}\ \bibnamefont {Small}},\ }\bibfield  {title}
  {\bibinfo {title} {Actin branching in the initiation and maintenance of
  lamellipodia},\ }\href {https://doi.org/10.1242/jcs.107623} {\bibfield
  {journal} {\bibinfo  {journal} {Journal of Cell Science}\ }\textbf {\bibinfo
  {volume} {125}},\ \bibinfo {pages} {2775 } (\bibinfo {year}
  {2012})}\BibitemShut {NoStop}%
\bibitem [{\citenamefont {Courson}\ and\ \citenamefont
  {Rock}(2010)}]{Courson:2010}%
  \BibitemOpen
  \bibfield  {author} {\bibinfo {author} {\bibfnamefont {D.~S.}\ \bibnamefont
  {Courson}}\ and\ \bibinfo {author} {\bibfnamefont {R.~S.}\ \bibnamefont
  {Rock}},\ }\bibfield  {title} {\bibinfo {title} {Actin cross-link assembly
  and disassembly mechanics for alpha-actinin and fascin},\ }\href
  {https://doi.org/10.1074/jbc.M110.123117} {\bibfield  {journal} {\bibinfo
  {journal} {J. Biol. Chem.}\ }\textbf {\bibinfo {volume} {285}},\ \bibinfo
  {pages} {26350} (\bibinfo {year} {2010})}\BibitemShut {NoStop}%
\end{thebibliography}

%

\begin{acknowledgements}
The authors thank Antoine Jégou and Guillaume Romet-Lemonne (Institut Jacques Monod, Paris, France) for providing proteins and Guy Tran Van Nhieu  (College de France, Paris, France) for hosting the purification of WASP-pWA and providing related tools. CVG, JH, OdR and ML were supported by ANR grant ANR-15-CE13-0004-03. MK was supported by the UpToParis program (European Union's Horizon 2020 research and innovation program under the Marie Skłodowska-Curie grant agreement No 754387). This work has received the support of Institut Pierre-Gilles de Gennes (Équipement d’Excellence, “Investissements d’avenir,” program ANR-10- EQPX-34). ML was supported by Marie Curie Integration Grant PCIG12-GA-2012-334053, “Investissements d’Avenir” LabEx PALM (ANR-10-LABX-0039-PALM), ANR-21-CE11-0004-02, ANR-22-ERCC-0004-01 and ANR-22-CE30-0024-01, as well as ERC Starting Grant 677532 and the Impulscience program of Fondation Bettencourt-Schueller. JH, OdR and ML’s groups belong to the CNRS consortium AQV.
\end{acknowledgements} 

\section*{Author contributions}
M.B. conceptualized the project, performed and analyzed the numerical simulations, wrote the original draft, reviewed and edited the manuscript.
C.V.G. performed and analyzed experiments.
M.K. performed and analyzed experiments.
L.K. performed preliminary numerical simulations.
G.F. helped supervise the numerical simulations.
J.H. conceptualized the project, designed and developed the experimental methodology, supervised the experiments, analyzed the experimental data, secured funding, reviewed and edited the manuscript.
O.d.R. conceptualized the project, designed and developed the experimental methodology, supervised the experiments, secured funding, reviewed and edited the manuscript.
M.L. conceptualized the project, supervised the numerical simulations, designed and performed the analytical theory and associated data analysis, secured funding, co-wrote, reviewed and edited the manuscript.

\section*{List of supplementary materials}
{\noindent}Supplementary text.\\
{\noindent}Supplementary figures S1 to S10.

\medskip
{\noindent}These materials are available as a .pdf ancillary file from this preprint's arXiv abstract page.

\end{document}